\documentclass[twocolumn,superscriptaddress,prl,showpacs,preprintnumbers,amsmath,amssymb]{revtex4}
\usepackage{amsmath}
\usepackage{amssymb}
\usepackage{dcolumn}
\usepackage{graphicx}
\usepackage{bm}

\begin{document}
\title{Reentrant and Isostructural Transitions in a Cluster-Crystal Former}
\author{Kai Zhang}
\affiliation{Department of Chemistry, Duke University, Durham, North
Carolina, 27708, USA}
\author{Patrick Charbonneau}
\email{patrick.charbonneau@duke.edu}
\affiliation{Department of Chemistry, Duke University, Durham, North
Carolina, 27708, USA}
\affiliation{Department of Physics, Duke University, Durham, North
Carolina, 27708, USA}
\author{Bianca M. Mladek}
\email{bmm32@cam.ac.uk}
\affiliation{Department of Chemistry, University of Cambridge,
Lensfield Road, CB2 1EW, UK}
\date{\today}

\begin{abstract}
We study the low-temperature behavior of a simple cluster-crystal
forming system through simulation. We find the phase diagram to be
hybrid between the Gaussian core model and the penetrable sphere model.
The
system additionally exhibits S-shaped doubly reentrant phase sequences
as well as critical isostructural transitions between crystals of
different average lattice site occupancy. Due to the possible
annihilation of lattice sites and accompanying clustering, the system
moreover shows an unusual softening upon compression.

\end{abstract}
 \maketitle


Since van der Waals and Kirkwood, we better appreciate the fundamental
role of harshly repulsive interactions in the organization of
matter~\cite{vanderwaals:1988}.  But what happens when harshness turns
into softness? Core softened potentials can exhibit microphase
separation~\cite{glaser:2007}, reentrant melting, and isostructural
phase transitions~\cite{young:1979}, as found in systems as diverse as
Cerium metal~\cite{young:1991}, star polymers~\cite{watzlawek:1999},
dipolar spheres~\cite{osterman:2007}, electron
bubbles~\cite{goerbig:2003,poplavskyy:2009}, and rotating Bose
gases~\cite{cooper:2008}. Even softer \emph{coreless} repulsive
interactions are also found in complex systems. Nonlinear fields can
form particle-like structures governed by soliton-like
interactions~\cite{aranson:1990,gomez:1995}; and for structures with
low fractal dimension, such as polymers~\cite{louis:2000},
dendrimers~\cite{likos:2002,mladek:2008b}, and
microgels~\cite{gottwald:2005}, the centers of mass can be immaterial
and thus overlap with only a finite free energy penalty. As a result,
materials governed by such interactions exhibit unusual phenomenology
compared to ``simple'' matter~\cite{likos:2001a}. Soft core models are
further used to study the difficult glass~\cite{klein:1994,ikeda:2010}
and classical ground state determination
problems~\cite{likos:2006,torquato:2008}, highlighting the broad
interest in the behavior of systems with soft potentials.

A certain universality permeates the thermodynamic assembly of systems
governed by bounded, purely repulsive interactions. Two
phenomenological categories have been identified from a mean-field
analysis. Systems with pair interactions whose Fourier components are
purely positive are expected to show reentrant melting, while those
with some negative Fourier components should cluster and freeze into
multiply occupied crystals (MOC) upon
compression~\cite{klein:1994,likos:1998,likos:2001,mladek:2006,mladek:2007,mladek:2008,lascaris:2010,ziherl:2010}. Crossing
the divide between the two categories can be realized via the tuneable
generalized exponential model of index $n$
(GEM-$n$)~\cite{mladek:2006}, whose pair potential for particles a
distance $r$ apart is
\begin{equation}
\phi(r) = \varepsilon
\exp\left[-(r/\sigma)^n\right],
\end{equation}
with $\varepsilon$ and $\sigma$ setting the units of energy and
length, respectively~\cite{footnote:1}.  Upon compression at low
temperatures, the $n=2$ Gaussian core model
(GCM)~\cite{stillinger:1976}, which has a purely positive Fourier
transform, forms cubic crystals that eventually
remelt~\cite{lang:2000,prestipino:2005}; while all GEM-$n$ of $n>2$
have some negative Fourier components and are thus predicted to form
MOC at high temperatures~\cite{likos:2001}, including the
$n\rightarrow\infty$ penetrable sphere model (PSM)
limit~\cite{likos:1998}.

For strong interactions (or effective low temperatures), which may be
the experimentally most relevant regime,
the phase behavior of MOC-forming systems is not understood. Various
plausible ordering scenarios upon compression are suggested by theory
and experiments: density functional theory predicts a continuous
increase in clustering~\cite{likos:2001,mladek:2006}; the PSM limit
presents a sequence of second-order phase transitions between crystals
of increasing occupancy~\cite{likos:1998}; and bubble solids alternate
liquid and crystal phases of increasing lattice
occupancy~\cite{goerbig:2003}. A cascade of pure first-order
isostructural solid-solid transitions between crystals of different
occupancy is also plausible, as theoretically predicted by
Ref.~\cite{neuhaus:2010}.  Such isostructural transitions should
terminate at critical points, because the cluster occupancy of
high-temperature MOC-formers increases continuously with
density~\cite{mladek:2006,mladek:2007}. A rare example of this type of
critical point is found in pure Cerium, where a pressure-induced
electronic promotion underlies the transition between two
isostructural solids with different lattice
\emph{spacing}~\cite{young:1991,lipp:2008}. Critical points involving
volume collapse have also been predicted for a variety of purely
classical, discontinuous model
potentials~\cite{young:1979,bolhuis:1997}. But the experimental
colloidal systems in which they could be observed have intrinsic size
polydispersity~\cite{buzzaccaro:2007} or smooth effective
interactions~\cite{likos:2001a} that inhibit phase separation. The
relatively broad lattice spacing of MOC and their smooth interaction
form suggest that crystal formation should be less sensitive to these
experimental constraints.

In this letter, we present a computational study of the
low-temperature behavior of the MOC-forming GEM-4. Its phase behavior
is shown to be a complex hybrid between the GCM and the PSM limits. We
furthermore find complex reentrant transitions and evidence for a
cascade of isostructural transitions.

\begin{figure*}
\includegraphics[width=\columnwidth]{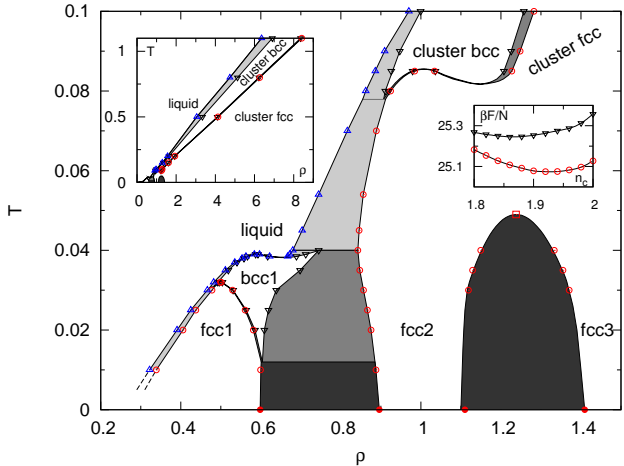}
\includegraphics[width=\columnwidth]{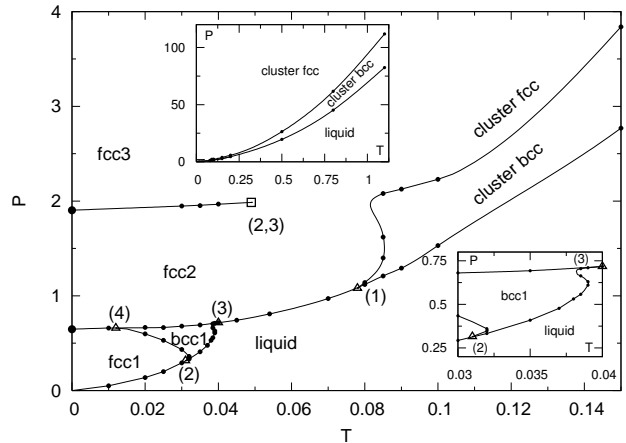}
\caption{(Color online) $T$-$\rho$ (left) and $P-T$ (right) low- and
  high-$T$ (left insets~\cite{mladek:2007}) simulation phase
  diagrams. The isostructural critical point ($\square$) is
  extrapolated from the law of rectilinear
  diameters~\cite{frenkel:2002} and the $T=0$ results ($\bullet$) come
  from phonon theory~\cite{neuhaus:2010}. Left: The coexistence
  regions (shaded) are delimited by simulation results for liquid
  ($\vartriangle$), bcc ($\triangledown$), and fcc ($\odot$) phases.
  The free energy per particle for the bcc ($\triangledown$) and fcc
  ($\odot$) phases at $T=0.03$ and $\rho=0.85$ shows that a fcc with
  $n_c^{\mathrm{eq}}=1.94(1)$ is the ground state (right inset).
  Right: The phase boundaries (solid lines) are guides for the eye
  that are consistent with the Gibbs-Duhem slopes (not shown) at the
  coexistence points ($\cdot$). The triple points ($\vartriangle$) are
  numbered. The right inset enlarges the SDR liquid-bcc1-liquid-bcc1
  sequence and the bcc1-fcc1-bcc1 crystal reentrance.}
\label{fig:trho}
\end{figure*}

\begin{figure}
\includegraphics[width=3.5in]{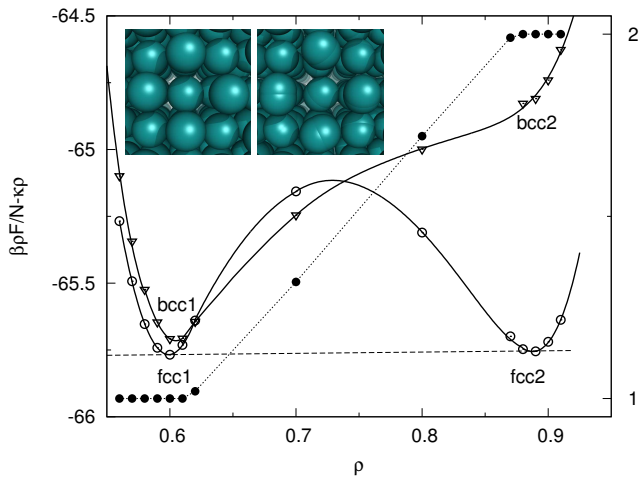}
\caption{Free energy curves of the stable fcc ($\odot$) and, for
  comparison, the metastable bcc ($\triangledown$) structures at
  $T=0.01$ with $\kappa=142.7$. Both curves show the van der Waals
  loop characteristic of a system whose limited size inhibits phase
  separation. The coexistence densities of the fcc1 (left inset)-fcc2
  (right inset) isostructrual transition are determined by the common
  tangent construction (dashed line). The equilibrium lattice
  occupancy $n_c^{\rm eq}$ ($\bullet$) plateaus near integer values
  for the thermodynamically stable phases.}
\label{fig:ncrho}
\end{figure}

We perform lattice Monte Carlo simulations~\cite{panagiotopoulos:2000}
of the GEM-4 model, whose high-temperature behavior was previously
determined~\cite{mladek:2007}, for $N=1000$-$5000$ particles at
constant $N$, volume $V$, and temperature $T$.  The pressure $P$ is
obtained from the virial and the Helmholtz free energy $F$ of the
different phases is calculated via thermodynamic integration. The
ideal gas reference~\cite{frenkel:2002} is complemented, for the
body-centered cubic (bcc) and face-centered cubic (fcc) crystal
phases, with a potential well centered around each of the $N_c$ lattice
sites~\cite{mladek:2008}.
This crystal reference, by allowing for the
characteristic multiple occupation of lattice sites and for particle
hopping between those sites, permits a reversible integration
path. For a fixed number density $\rho$, the average lattice site
occupancy $n_c = N/N_c$ at equilibrium is identified for every state
point by simulating crystals at various fixed $n_c$ then minimizing
the resulting constrained free energy $F(n_c)$, i.e., identifying the
loci $F(n_{c}^{\mathrm{eq}})$ such that (Fig.~\ref{fig:trho})
\begin{equation}
\left[\frac{\partial F(n_c)}{\partial
    n_c}\right]_{\rho,T,n_c=n_c^{\mathrm{eq}}}=0,
\end{equation}
similarly to the scheme employed in Ref.~\cite{zhang:2010} and that
used for GEM-4's high-$T$ phase diagram determination.  In the latter,
$n_c$ was tuned until the (unphysical) field conjugate to $n_c$ had
vanished~\cite{mladek:2007,swope:1992}, which allows for a
gradient-based minimization of $F$, but relies on an additional
independent calculation of the chemical potential $\mu$. 
This approach breaks down at low $T$, where $\mu$ cannot be
efficiently resolved by Widom's particle
insertion~\cite{frenkel:2002}. Phase coexistence is then determined
through common tangent construction of the free energy data. A linear
transformation $\beta\tilde{F}\rho/N= \beta F\rho/N - \kappa\rho$ with
inverse temperature $\beta$ and arbitrary unitless parameter $\kappa$
enhances the visibility of the coexistence regime
(Fig.~\ref{fig:ncrho}).

We present two projections of the the low-$T$ phase diagram in Fig.~\ref{fig:trho}. As
anticipated from the high-$T$ extrapolation, the cluster bcc phase
vanishes at a triple point $T_{\mathrm{t}}^{(1)}=0.078(1)$. But
surprisingly the transition is preceded by a S-shaped doubly reentrant
(SDR) crystal phase sequence, where both a maximum and a minimum in
the coexistence curve are observed.  Below $T_{\mathrm{t}}^{(1)}$ the
phase diagram is surprisingly rich. GCM-like phase
behavior~\cite{stillinger:1976,lang:2000,prestipino:2005} is followed
by an unexpectedly complex clustering regime. Because the tail of the
GEM-4 decays faster than any inverse power, the liquid freezes into a
single-occupancy fcc (fcc1) that reaches vanishingly small densities
at low $T$, in agreement with predictions from genetic
algorithms~\cite{mladek:2007b,mladek:2007c} and phonon
theory~\cite{neuhaus:2010}.  This fcc1 phase gives way to a
single-occupancy bcc (bcc1) phase at a second triple point
$T_{\mathrm{t}}^{(2)}=0.031(1)$. For a narrow temperature range above
$T_{\mathrm{t}}^{(2)}$, a bcc1 wedge between the liquid and the fcc1
phase leads to reentrant crystallization of bcc1 upon compression. The
maximum freezing temperature for bcc1 at $T=0.039(1)$ leads to a
second SDR sequence similar to that observed in the hard core/soft shoulder model~\cite{young:1979,young:1991}.  By contrast to the GCM's
reentrant melting
behavior~\cite{stillinger:1976,lang:2000,prestipino:2005}, here the
liquid reentrance section of the SDR sequence spreads only over a
finite density regime $0.59\lesssim\rho\lesssim0.68$ and over a much
smaller temperature range $0.0385\lesssim T\lesssim0.039$. The
intermediate nature of the GEM-4 suggests that this reentrance might
become more pronounced as the GCM is approached, i.e.,~$n\rightarrow
2^+$, and should disappear before the PSM limit $n \rightarrow
\infty$, where reentrance is not expected~\cite{likos:1998}. The
connection between the high and low-$T$ regimes occurs through a third
triple point $T_{\mathrm{t}}^{(3)}= 0.040(1)$, at which bcc1 vanishes.
A prior, coarser study of the liquid-crystal transition in this regime
missed both the presence of fcc1 and of the reentrant
melting~\cite{fragner:2007}. It also inaccurately assigned the unusual
shape of the liquid-crystal coexistence curve to the onset of
clustering, while it is rather caused by the reentrant
melting.

\begin{figure}
\begin{center}
\includegraphics[width=3.5in]{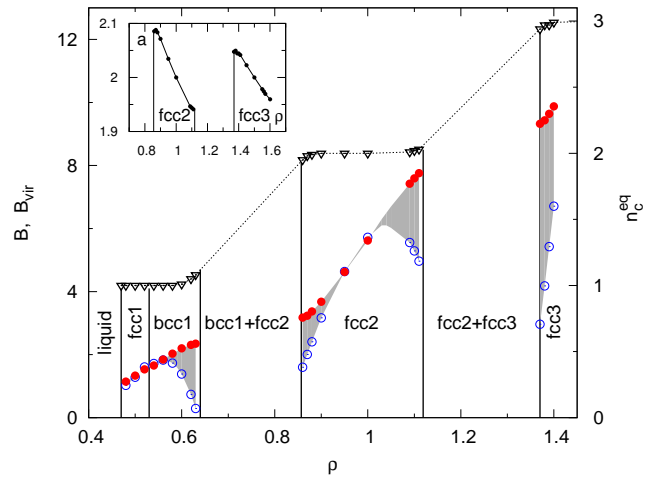}
\caption{(Color online) Isothermal ($T=0.03$) bulk modulus $B$
  ($\odot$), its virial contribution ($\bullet$) and softening
  correction (shaded), and $n^{\mathrm{eq}}_c$
  ($\triangledown$)~\cite{footnote:2}. The solid vertical lines
  indicate phase boundaries. The lattice constant $a$ changes
  discontinuously as fcc2$\rightarrow$fcc3 (inset).}
\label{fig:B}
\end{center}
\end{figure}

Clustering does indeed influence the phase diagram topology,
but at densities further away from the liquid phase. At low $T$ the
nature of clustering is unlike what is seen at higher $T$, where $n_c$
changes linearly with $\rho$ resulting in a nearly density independent
lattice constant~\cite{mladek:2007}. Here, $n_c$ is quasi-quantized,
and at very low $T$ the lattice constant $a$ changes discontinuously
through isostructural transitions between fcc lattices of nearly
perfect integer mean occupancy $n_c \leftrightarrow n_c+1$
(Figs.~\ref{fig:ncrho}-\ref{fig:B}). The first occurrence of these
transitions, fcc1 $\leftrightarrow$ fcc2, is partially interrupted by
the bcc1 phase, down to the fourth triple point
$T_{\mathrm{t}}^{(4)}=0.012(1)$ (Fig.~\ref{fig:trho}). But at higher
densities, the fcc2 $\leftrightarrow$ fcc3 coexistence is fully
developed. No other liquid or crystal phases are found to interfere
and genetic algorithm results corroborate that no other crystal
symmetry should be stable in this density
regime~\cite{mladek:2007b,mladek:2007c}. It is at the moment
computationally difficult to go beyond fcc3, but both a zero
temperature treatment paired with phonon theory~\cite{neuhaus:2010}
and a simple mean-field cell theory predict a cascade of $n_c
\leftrightarrow n_c+1$ isostructural transitions to carry on \emph{ad
  infinitum}, slightly broadening the coexistence regime between two
integer occupancies. As argued above, the topology of the phase
diagram demands that each uninterrupted isotructural transition
terminates at a critical point, the first one of which is found at
$T_c^{(2,3)}=0.049(3)$. Hopping between lattice sites should depress
$T_c^{(n_c,n_c+1)}$ as $n_c$ increases, and mean-field critical
universality is generally expected~\cite{chou:1996,likos:2001}.  The
series of first-order isostructural transitions contrasts with the
continuous second-order clustering transitions for the PSM predicted
by cell theory~\cite{likos:1998}. This last observation suggests that
the behavior of the PSM might be singular, but further studies are
necessary to clarify the GEM-$n$ family phase behavior as
$n\rightarrow \infty$.

One of the key material properties of MOC is the presence of two
distinct microscopic mechanisms for responding to compression. Like
any other crystal, MOC can affinely reduce their lattice constant, but
additionally they can eliminate lattice sites by increasing the mean
lattice occupancy. We can decompose the bulk modulus
\begin{equation}
B\equiv V\left(\frac{\partial^2 F}{\partial
  V^2}\right)_{N,T}=B_{\mathrm{vir}}-B_{\mathrm{corr}}
\end{equation}
into a constant $n_c$ virial contribution $B_{\mathrm{vir}}$ and a
``softening'' correction $B_{\mathrm{corr}}$, which map directly to
the two microscopic mechanisms~\cite{rowlinson:1959,mladek:2007}.  At
high $T$, $B_{\mathrm{corr}}$ can be as high as half the virial
component~\cite{mladek:2007}, but in low $T$ crystals, the
quasi-quantized jumps in $n_c$ lead to a significantly different
mechanical behavior. Away from the coexistence regions, where $n_c$ is
nearly constant, the system responds only affinely to isothermal
compression and the virial contribution to the bulk modulus captures
the full response of the system, i.e., $B_{\mathrm{corr}}\sim 0$,
(Fig.~\ref{fig:B})~\cite{footnote:2}. But in the softening regions
that precede and follow the phase transitions the quantization is
imperfect, and $B_{\mathrm{corr}}\ne 0$. Near the bcc1-fcc2
transition, for instance, $B_{\mathrm{corr}}$ is nearly equal to the
virial contribution, which means that the system exerts almost no
resistance to compression. This very rapid change in mechanical
properties with compression is uncommon, and may lead to novel
material behavior. The different physical natures of the virial and
softening contributions indeed suggest a separation of time scales for
their microscopic relaxation, with slow particle redistributions
contrasted by fast affine deformations. Hardening or softening of the
material upon compression might thus depend on the deformation rate.

We have presented the intriguing low-temperature phase behavior of the
MOC-forming GEM-4 through a numerical method specially designed for
this class of systems.  The complexity of the phase behavior is
particularly noteworthy considering the simplicity of the model, which
is free of competing length scales. Experimental soft matter
realizations of MOC are still lacking, but large-scale,
monomer-resolved simulations of amphiphilic cluster-forming dendrimers
are currently under way~\cite{likos:priv}. Importantly, the approach
outlined in the present work should be directly applicable to
phenomena of reversible cluster formation in other branches of
physics. Examples are: the structures formed by the soft
solitons~\cite{aranson:1990,gomez:1995}, the quasi-2D electron bubbles
in the quantum-Hall regime~\cite{goerbig:2003,poplavskyy:2009}, and
the predicted clustering of vortex lines in rotating Bose
gases~\cite{cooper:2008}.

\begin{acknowledgments}
We thank D. Frenkel, O. Poplavskyy, N. Cooper (Cambridge), and
C. N. Likos (Vienna) for helpful discussions, and A. Dawid (Grenoble)
for careful reading of the manuscript. KZ and PC acknowledge ORAU and
Duke startup funding. BMM acknowledges EU funding via
FP7-PEOPLE-IEF-2008 No. 236663.
\end{acknowledgments}


\begin{thebibliography}{46}
\bibitem[{foo({\natexlab{b}})}]{footnote:2}
\bibinfo{note}{The bulk modulus is obtained by differentiating empirical fits
  to the free energy results. Though this approach captures the mechanical
  behavior rather well, some numerical errors remain since the softening
  correction must clearly be positive.}
\end{thebibliography}

\end{document}